\documentclass{article}

\title{Ultra-High Energy Cosmic Rays and Absolute Reference Frame
  defined by External Field}

\author{Humitaka SATO\\
Department of Physics, Konan University\\
  Okamoto, Kobe 658-8501, JAPAN \\ 
E-mail: satoh@konan-u.ac.jp
}
\begin{document}
\maketitle

\begin{abstract}

High-energy end of the cosmic-ray spectrum has provided us to check a validity of the Lorentz Invariance and the Relativity principle, through the observation of the so-called GZK cut-off. It is claimed in this report that the comoving reference frame in the expanding universe might define the preferable inertia frame, in contradiction to the  relativity principle. If the present universe has been permeated by  tensor  fields in a manner like it has been done by Higgs scalar  field , the limiting particle velocity of each species  splits to different values depending on the coupling coefficients to these external fields.
\end{abstract}

\section{Historical Introduction }
Energy spectrum of the cosmic rays extends by a power-law over more than ten decimal, decreasing in a power-law like $E^{-\gamma}$ with energy $E$ and $\gamma \sim 2.5$. Then a natural question is whether  the high-energy end in the energy spectrum does exist or not. In 1966, a very clear-cut prediction was presented, which introduced a definite upper-limit in the power-low energy spectrum, which is called now as GZK(Greisen-Zatepin-Kuzmin) cut-off.\cite{gz}

This cut-off prediction was invoked by the discovery of "3K radiation" in 1965, which is now called as CMB(cosmic microwave background). Although CMB was observed just on the earth,  CMB was supposed to fill up in the whole cosmic space uniformly, even in the extra-galactic space, as the relics of "hot" big-bang. Therefore it became crucial to check the presence of CMB in the extra-galactic space, in order to settle  the big debates between the steady state cosmology and Gamow's hot big-bang cosmology. As an advocator of the steady-state cosmology, Fred Hoyle tried hard to present two types of counter arguments, one was about exotic interstellar dusts which masks the extragalactic view in this wave-band and another one was how much degree the high-energy cosmic radiations($\gamma$-rays,X-rays,electron-positron,etc) are masked by the extra-galactic presence of CMB. Hoyle's motivation of the latter argument was to point out a contradiction of the hot big-bang cosmology but thi!
s argument had created a rich implication of CMB toward the high-energy cosmic radiations, including the GZK cut-off.

Since then, the GZK cut-off energy of about $10^{20}$eV became an experimental target for the cosmic-ray physicist. Observation of the EAS(extensive air shower) started also in Japan and EAS-group led by K. Suga constructed the array of detectors in the suburb of Tokyo, a dense array in the site of the research institute and several remote stations at the sites of elementary school, the city office and so on. It was the autumn of 1971 when they announced that their detector had catched  a huge EAS with energy over GZK cut-off in 1970. In following February, the workshop was organized in order to discuss this puzzling EAS event. 

At the workshop, I gave a talk by the title "Very high-energy cosmic-rays and the limitation of relativity principle".\cite{s2} If the high-energy end does not exist contrary to the GZK cut-off prediction, we could enumerate three possible ways of resolution, 1) "3K"radiation is local, 2)source of such cosmic ray is local(within mean-free-path), 3) cosmic ray is not proton but some exotic primary. At the workshop of 1972, I added the fourth possibility 4) violation of relativity principle. Later the paper was written by the title "Ultra-high Energy cosmic rays, Hot universe and Absolute reference frame".\cite{72} 

Although the energy estimation of this 1970-event was not accurate enough to claim the existence of super-GZK cut-off cosmic rays, this event promoted very much the effort toward a construction of bigger array in Akeno. This new big array, AGASA, finally presented more assured experimental evidences of super-GZK cut-off after 1997.   Experimental data suggesting super-GZK cosmic rays given by AGASA\cite{ag} as well as FlysEyes gave a great impact towards the bigger new observational projects such as Auger, EUSO\cite{ta} , and others. 

In such trend of research, an implication of the super-GZK cosmic ray has been discussed widely. In different from the situation in 1972, the first possible way(local "3K") has been eliminated and other three possibilities have been discussed; a) exotic local source such as cosmic string, mini black hole,etc.(so-called top-down scenario), b)exotic primaries such as neutrinos, neutrino with Z-burst in Galactic halo, etc, and finally c)violation of Lorentz invariance. 

Even for the last possibility, there are variety of arguments.\cite{{98},{a8},{00},{01}} In this report, a specific toy model of violation of Lorentz invariance is proposed and an extension of Lorentz invariance with non-unique limiting velocities is discussed.

\section{Comoving Frame in the Expanding Universe and Relativity Principle}

In the expanding universe, we can clearly identify  preferential inertia frames: (1)rest frame of baryon matter, (2)rest frame of astronomical objects, (3)frame in which CMB is isotropic, (4) frame in which the Hubble flow is observed isotropic. Furthermore, these four frames are approximately identical within a relative velocity difference of several hundreds km/sec. These inertia frames have a concrete physical effect when we understand the structure formation in the expanding universe.\cite{01} 

According to recent theoretical view on the early universe, these cosmological frames are considered to have the same physical origin; spontaneous selection of the inertia frame in  which the primordial black body radiation is isotropic via a reheating at Inflation. But even in the vacuum universe without material substance, the creation of the expanding universe itself is  the browken state of  Lorentz invariance. That is a formation of comoving frame  perpendicular to the time direction. We call this cosmological and comoving frame as C-frame.

In spite of a lucid presence of the C-frame, however, the Lorentz invariance  is supposed to hold in any local physical phenomena. The relativity principle does not respect this lucid presence. Whatever lucid this presence is, it  has no physical effect. That is the spirit of the relativity since Galileo. In the derivation of GZK cut-off, the relativity principle is used as usual but its situation is very special because the Lorentz factor relative to the C-frame is as large as $\gamma \sim 10^{11}$, which is far beyond the Lorentz factor in the particles  the accelerators of about $\gamma \sim 10^5$.

Here we should not confuse the two meanings of "high energy".
One is an invariant energy(or center of mass energy) defined such as ,
$$p^\mu p_\mu=E^2-P^2=Q^2$$
,where $p^\mu$ is total four momentum of the system.
Another one is  energy relative to a specific reference frame and it will be defined in the following manner as
$$ N^\mu p_\mu=1\cdot E-0\cdot P=E$$
, where $N^\mu$ is a four vector specifing the frame. For the C-frame, the component is given as $N^\mu(1,0,0,0)$ in the C-frame. The Relativity principle claims that the cross section of collision, $\sigma$, does depend solely on $Q$ but does  not depend on $N^\mu p_\mu$ , such as $\sigma(Q)$ but not as $\sigma(Q, N^\mu p_\mu)$.  In our early paper\cite{72}, the cut-off function in the momentum space was assumed to depend on $N^\mu p_\mu$ and the cross section involved to the GZK was altered not to give the cut-off of the spectrum. 

In the discussion of GZK cut-off, $Q$ is $\sim 10^{8.5}$eV, which is rather low energy in  high-energy physics, but, $N^\mu p_\mu \sim 10^{20}$eV is extraordinarily large even in high-energy physics. The uniqueness of the GZK cut-off lies on the largeness of $N^\mu p_\mu$, but not on the so-called energy frontier of the high-energy physics, e.g., Energy frontier for supersymmetry, GUT, Planck scale, etc., those are talking about large $Q$ but not on the largeness of $N^\mu p_\mu$.  

\section{A Toy Model of Lorentz-Invariance Violation}

Consider the following Lagrangian for a Dirac particle A,

$$ L_A={i \over 2}\bar{\psi}\gamma_\mu \partial^\mu \psi-\alpha_A \phi \bar{\psi}\psi+{i \over 2}g_A F_{\mu \nu}\bar{\psi}\gamma^\mu\partial^\nu \psi,$$

where $\psi$ is the Dirac field of A, $\phi$ is Higgs scalar field with coupling coefficient $\alpha_A$ and $F_{\mu \nu}$ is a tensor field with coupling coefficient $g_A$.The first term in the right hand side is kinetic term and the second one is the Yukawa coupling term which creates mass by Higgs mechanism.  In this Lagrangian, the dynamical parts of $\phi$ and $F^{\mu \nu}$  has been omitted and $\phi$ and  $F^{\mu \nu}$ are both taken as an external field. They are un-removable given field in the present state of universe. Non-zero value of $<\phi>$ gives the mass, $m_A=\alpha_A <\phi>$, to this Dirac particle.

Next we assume that some component of the tensor field has got some non-zero value as followings,

$$<F^{00}>=B\neq 0~~ {\rm and} <F^{\mu\nu}>=0~~ {\rm for~~other~~components}. $$$B$ is supposed to be constant in space and time but can be slowly changing with cosmological spacetime scale. 
Then the dispersion relation for plain wave is given as\cite{bt}

$$p^\mu p_\mu -m_A^2c^2=-2g_AB(E/c)^2$$
, where only the first order terms of B has been retained and the higher term of B has been neglected.

This relation is rewritten by denoting the three momentum as $p$ as
$$(1+g_AB)(E/c)^2=p^2+m_A^2c^2,$$
where $c$ is the universal constant introduced at the definition of the spacetime length by space length and time length.

Renormalizing the velocity and mass as followings
$$ c_A^2={ c^2 \over 1+g_A B}~~ {\rm and}~~m_{AB}^2=(1+g_AB)m_A^2 ,$$
the conventional energy-momentum relation is resumed

$$ E^2=p^2 c_A^2 + m_{AB}^2c_A^4.$$
but now $c_A$ is depending on particle species through $g_A$, that is, the limiting velocity, velocity in the limit of $E \rightarrow \infty$, is depending on the particle species.

Here we remark some difference between the Higgs scalar $\phi$ and the tensor external field $F^{\mu \nu}$. Different from a scalar field , we have adopted the C-frame as the preferential frame and the above energy-momentum relation holds only in the C-frame. If we modified the Lorentz transformation with psudo-Lorentz factor
$$ \gamma_A={1 \over \sqrt{1-\left({ v \over c_A}\right)^2}}~~{\rm instead~~of~~} \gamma={1 \over \sqrt{1-\left({ v \over c}\right)^2}},$$
the above relation keeps its form. However the Lorentz invariance apparently breaks down if we consider a system consisting of pariticles of different species. 

The perturbative super string theory has suggested an existence of various hidden fields such as the above tensor field.\cite{ks} 
If we assume a vector field $A_\mu$ in stead of $F_{\mu \nu}$ as the external field, the Lagrangian is written,\cite{kl}
$$ L_A={i \over 2}\bar{\psi}\gamma_\mu \partial^\mu \psi-m_A \bar{\psi}\psi-f_AV_\mu \bar{\psi}\gamma^\mu  \psi.$$
, where the Higgs term is now rewritten by the mass term.
Here we assume 
$$ <V_0>=V\neq 0 ~~{\rm and}~~<V_\mu=0>~~{\rm for~~all~~other~~ components}$$
and the the dispersion relation becomes like
$$E^2-p^2c^2-m_A^2c^4=-2f_AVE.$$
If we define as
$$c_A(E)={ c \over 1+{f_AV \over E}},~~ m_{AV}^2=(1+f_AV/E)^2 [m_A^2+(f_A V)^2/c^4],$$
the above dispersion relation resume  a pseudo-conventional form like
$$E^2=p^2c_A(E)^2+m_{AV}^2c_A(E)^4.$$
$c_A(E)$ has anomaly in the limit of $E \rightarrow 0$ but this limit would need a quantum mechanical correction. The violation of Lorentz invariance would dominate in the vector case similar to the scalar or Higgs case. Then the tensor case is necessary as the toy model which exhibits the violation of Lorentz invariance in the limit of large $\gamma$

\section{Boost Particle-Transformation in the External Field}

The above argument can be discussed from a different viewpoint. We can consider two types of transformation, boost particle-transformation and the Lorentz transformation.\cite{ck} The Lorentz transformation is just a change of reference frame for the description of the same phenomena and is sometime called "passive" transformation. The boost particle-transformation  is "active" transformation, where particle's energy-momentum are changed actually. Relativity principle claims that the boosted state and the original state seen from the transformed reference frame are identical. For the system of particles, this is trivial and the classification into "Boost" and "Lorentz" has no particular meaning.

However some complication comes in when we consider the system consisting of  particles and external given field. In the Lorentz transformation, both the  particle's energy-momentum and the components of the external field are transformed. Therefore the relative relation between particle and external field does not changed. In the boost particle-transformation, however, particle's energy-momentum are transformed but the field configuration is kept unchanged. Therefore two states of the particles relative to the field are different.    
In this way, the actively boosted state of particle is not identical with the passively Lorentz transformed state having the same particle state but different field configuration. Thus we call this situation as an "apparent" violation of Lorentz invariance  but it is in fact a misconduct of the Lorentz transformation.

What we have done in the previous section is something like this. In the actual universe, the external fields like $F^{\mu \nu}$ are totally unknown to us upto now and "misconduct" of application of the Lorentz transformation could happen. Conversely we also say that the apparent violation implies a finding of the hidden  external fields.

\section{Eigen State of the Limiting Velocity and GZK cut-off}
Without touching on the origin of various limiting velocity, we can rise a question how much degree the universality of limiting velocity has been checked by direct experiment. The assumption of non-equality of the limiting velocity of a charged particle and light velocity  is equivalent to the  introduction of the Lorentz non-invariant term of the electromagnetic field into the Lagrangian.\cite{gd} In general, this is  true for any  non-universal assumptions of the limiting velocity.\cite{c9}

Coleman and Glashow also discussed this assumption, firstly in order to explain the neutrino oscillation.\cite{c7} They also pointed out that the high-energy phenomena might disclose  an apparent degeneracy of limiting velocity and reveal  a splitting into a fine structure. They called various limiting velocity as eigen state of velocity. They have shown also that this modification does not hurt the standard theory of interaction based on the gauge field theory.\cite{c9}
The discussion in the section 3 is concerned the origin of such an ad hoc assumption of the eigen state of limiting velocity.

If we introduce the particle species dependent $c_A$, the GZK cut-off discussion could be  modified very much. By the head-on collision between the cosmic-ray proton and the CMB photon, $\Delta$ particle is produced if the following condition is satisfied.\cite{bt}
$$(E_p+E_\gamma)^2-(p_p+p_\gamma)^2c_\Delta^2 >m_\Delta^2 c_\Delta^4,$$
while the proton obeys to $E_p^2=p_p^2c_p^2+m_p^2c_p^4$. In the situation of $E_p \gg m_p c_p^2$ and $\vert c_\Delta-c_p \vert\ll c_p$, the  condition becomes as followings
$$-{c_\Delta-c_p  \over c_p}E_p^2+2E_pE_\gamma >{m_\pi^2c^4 \over 2}$$
In the conventional case, $c_\Delta-c_p =0$ and the threshold energy is obtained $E_p>m_\pi^2 c^4/4 E_\gamma$. 

If $(c_\Delta-c_p)\neq 0$, the above equation gives a quite different result; the cut-off  disappears for $(c_\Delta-c_p)>0$ and the cut-off energy  decreases compared with the GZK cut-off for $(c_\Delta - c_p)<0$. For example, the above equation does not have solution if
$${c_\Delta-c_p \over c_p} >2\left({E_\gamma \over m_\pi c^2}\right)^2\sim 10^{-22},$$
the cut-off does not exist.

On the other hand, for $(c_\Delta-c_p)<0$, the cut-off energy is modified as
$$E_{GZK}\left[1- {\vert c_\Delta-c_p\vert \over 2c_p}\left({m_\pi c^2 \over E_\gamma}\right)^2\right]~~{\rm for}~~{\vert c_\Delta-c_p\vert \over 2c_p}\left({m_\pi c^2 \over E_\gamma}\right)^2 < 1$$
and 
$$ \sqrt{c_p \over 2\vert c_\Delta-c_p \vert} m_\pi c^2~~{\rm for}~~{\vert c_\Delta-c_p \vert \over c_p}m_\pi^2 c^4 \gg E_\gamma^2$$

\section{Paradigm of Spontaneous Symmetry Breakdown }
One of the achievement of the 20-century Physics was discovery of various symmetry hidden deep in the diversity of superficial phenomena: we can point out many  symmetries such as rotational and boost  symmetry of 3-space, past-future symmetry in mechanics, duality symmetry between electro- and magneto-fields, Lorentz symmetry of spacetime, discrete symmetry in atomic structure of solid, particle-antiparticle symmetry,  isospin symmetry of nuclear force, chiral symmetry, "eight-fold symmetry", super-symmetry, colour symmetry and so on. Particularly, in the late of 1970's, theory of  fundamental interactions among elementary particles was formulated  into the unified-gauge-theory,  based on internal or local symmetry hidden in electro-weak and strong interactions among quarks and leptons. 

This unification of the fundamental interaction was accomplished, however, by one extra idea called "spontaneous symmetry breakdown(SSB)", which is schematically written as
$${\rm [observed~~ law]=[symmetric~~ law]x[SSB]}.$$
That is,  the symmetric law itself is not realized in this universe because the universe is not empty but the  external field called Higgs field has  permeated by . The most essential difference of the Higgs field from a conventional field is that it is un-removable from the universe. Then the genuine symmetric law looses its chance to exhibit  its original form in this universe. 

This SSB has introduced a new ingredient about the concept of  physics law, that is, the physics law itself is symmetric but our actual universe is not in a state of exact symmetry. This may be re-phrased also as followings; physics law is universal but our universe is not universal entity, or, physics law itself does not exhibit its original form in our universe where we live in. We call this kind of idea as the SSB paradigm.\cite{00} 

In fact, some symmetries are not exact but show a tiny breakdown, like in case of CP-asymmetry. The actual composition of cosmic matter does not obey the particle-antiparticle symmetry in spite of CPT-symmetry in physics law itself. Following these considerations, we are tempted to think that any symmetry might be not exact in this actual universe, which has come into an existence through various spontaneous selections of non-universal parameters.  

Lorentz invariance claims that there is no preferential inertia frame; that is the central dogma of relativity  principle. However, in our universe filled with the CMB and cosmic matter, we can clearly identify the preferential frame, which we have called the C-frame. In the inflationay scenario, CMB is supposed to be created in association with some SSB of the vacuum state of quantum field theory. Some features of the particle interaction in this universe is supposed to have inherited the parameters chosen by a dynamical process  of this SSB.  Furthermore, the SSB paradigm is now extended to the creation of spacetime from higher dimensional space through a dynamical process similar to SSB.  Thus we can speculate also the exact Lorentz symmetry might have been  violated  dynamically in "our universe", that is spontaneous breaking of Lorentz symmetry.\cite{ks}

Lorentz symmetry, however, has been built in all fundamental concepts of  modern physics, such as  Dirac field, spin, renormalization group of quantum field theory, and so on. Therefore, the violation of this symmetry can not be introduced so easily. One of the outcomes of the relativity principle is the equivalence of all inertia frame. However this equivalence has not been directly proved so much.\cite{w7} Only the accelerator experiments has proved this equivalence  up to some Lorentz factor of $\gamma_{\rm acce} \sim 10^5$. In this respect, the GZK cut-off has an unique status for the experimental verification of the equivalence of all inertia frames and the validity limit may be extended up to $\gamma_{\rm GZK} \sim 10^{11}$. Following to the SSB paradigm, this verification has coupled with the universality of the limiting velocity. And if there  were not the GZK cut-off, that may imply a finding of a un-removable hidden external field of tensor type.   
The SSB paradigm anyway describes our universe as "un-universal" universe.

\end{document}